\documentclass[a4paper,twocolumn,
english,aps,pra,floatfix,showpacs]{revtex4}
\usepackage[latin1]{inputenc}
\usepackage{amsmath}
\usepackage{babel}
\usepackage{graphics}
\usepackage{amssymb}

\begin{document}

\newcommand{\dt}{\Delta\tau}
\newcommand{\al}{\alpha}
\newcommand{\ep}{\varepsilon}
\newcommand{\ave}[1]{\langle #1\rangle}
\newcommand{\have}[1]{\langle #1\rangle_{\{s\}}}
\newcommand{\bave}[1]{\big\langle #1\big\rangle}
\newcommand{\Bave}[1]{\Big\langle #1\Big\rangle}
\newcommand{\dave}[1]{\langle\langle #1\rangle\rangle}
\newcommand{\bigdave}[1]{\big\langle\big\langle #1\big\rangle\big\rangle}
\newcommand{\Bigdave}[1]{\Big\langle\Big\langle #1\Big\rangle\Big\rangle}
\newcommand{\braket}[2]{\langle #1|#2\rangle}
\newcommand{\up}{\uparrow}
\newcommand{\dn}{\downarrow}
\newcommand{\bb}{\mathsf{B}}
\newcommand{\ctr}{{\text{\Large${\mathcal T}r$}}}
\newcommand{\sctr}{{\mathcal{T}}\!r \,}
\newcommand{\btr}{\underset{\{s\}}{\text{\Large\rm Tr}}}
\newcommand{\lvec}[1]{\mathbf{#1}}
\newcommand{\gt}{\tilde{g}}
\newcommand{\ggt}{\tilde{G}}
\newcommand{\jpsj}{J.\ Phys.\ Soc.\ Japan\ }

\def\jap{J.\ Appl.\ Phys.\ }
\def\jpcm{J.\ Phys.: Condens.\ Matt.\ }
\def\phe{Physica E\ }
\def\cH{{\cal H}}
\def\m{{\cal M}}

\title{Magnetic-field effects in defect-controlled ferromagnetic 
Ga$_{1-x}$Mn$_x$As semiconductors} 

\author{Raimundo R.\ \surname{dos Santos}}


\affiliation{Instituto de F\'\i sica, Universidade Federal do
Rio de Janeiro, Caixa Postal 68528, 21945-970
Rio de Janeiro RJ, Brazil}

\author{Luiz E.\ \surname{Oliveira}}

\affiliation{Instituto de F\'{\i}sica, Unicamp, C.P. 6165,
13083-970 Campinas SP, Brazil}

\author{Jos\'e d'Albuquerque e \surname{Castro}}
\affiliation{Instituto de F\'\i sica, Universidade Federal do
Rio de Janeiro, Caixa Postal 68528, 21945-970
Rio de Janeiro RJ, Brazil}

\date{\today}

\begin{abstract}
We have studied the magnetic-field and concentration dependences of the 
magnetizations of the hole and Mn subsystems in diluted ferromagnetic 
semiconductor Ga$_{1-x}$Mn$_x$As. 
A mean-field approximation to the hole-mediated interaction is used,
in which the hole concentration $p(x)$ is parametrized in 
terms of a fitting (of the hole effective mass and hole/local moment coupling)
to experimental data on the $T_c$ critical temperature.  The dependence of the
magnetizations with $x$, for a given temperature, presents a sharply peaked
structure, with maxima increasing with applied magnetic field, which indicates
that application to diluted-magnetic-semiconductor devices would require
quality-control of the Mn-doping composition.  We also compare various 
experimental data for $T_c(x)$ and $p(x)$ on different Ga$_{1-x}$Mn$_x$As 
samples and stress the need of further detailed experimental work 
to assure that the experimental measurements are reproducible.
\end{abstract}
\maketitle

Diluted magnetic semiconductors (DMS) have become one of the most promising
classes of materials for spintronics applications.  This is mainly due to the
possibility of manipulating both the charge and spin degrees of freedom of
electrons or holes to process and store information in magnetic materials.  The
discovery of hole-induced ferromagnetism in p-type (In,Mn)As systems
\cite{Ohno92} was followed by the successful growth of ferromagnetic (Ga,Mn)As
alloys \cite{Ohno96}.  Interest in the understanding of the physics in these
materials has boosted due to the fact that ferromagnetic III-V alloys may be
readily combined into semiconductor heterostruture systems, opening up a range
of applications of optoelectronic devices through the combination of quantum and
magnetic phenomena in these materials.  However, several issues in relation to
these systems need to be elucidated before full-scale applications can be
efficiently implemented.  For instance, a light-emitting device based on III-V
heterostructures has been proposed \cite{YOhno99}, which relies on the injection
of holes from a (Ga,Mn)As layer in the presence of a magnetic field.  Therefore,
the magnetic response of the hole subsystem should be known to some detail,
which, in turn, must reflect the dependence of the hole concentration with the
Mn composition.  The latter is still a challenging problem:  While in principle
each Mn atom should provide one hole, leading to a density of holes, $p$, equal
to that of the magnetic ions, early experimental data already indicated that $p$
is only a 15 to 30\% fraction of that of magnetic
ions\cite{Matsukura98,Ohno-tc}.  We have recently addressed this issue
\cite{dSOC} through a mean-field approximation to a Hamiltonian incorporating
the hole-mediated mechanism \cite{Dietl97} and found that the hole concentration
displays a non-monotonic behavior with a maximum near $x=0.05$, within the
insulating phase.  In spite of its simplicity, this mean-field framework should
provide a qualitative description of the response to a magnetic field.  With
this in mind, here we obtain the magnetization of both the hole `gas' and of the
Mn subsystem, as functions of the magnetic field and of the Mn composition.

\begin{figure}[h]
{\centering\resizebox*{2.5in}{!}{\includegraphics*{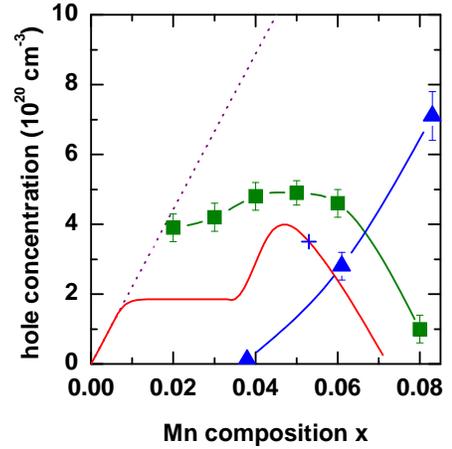}}}
\caption{\label{fig1} Hole concentration as a function of Mn composition in
Ga$_{1-x}$Mn$_x$As alloys:  
the full curve is the theoretical result obtained in 
Ref.\ {\protect\cite{dSOC}}, 
the cross corresponds to the experimental datum quoted in 
Ref.\ {\protect\cite{Ohno-tc}},
the full squares are the experimental data by Edmonds 
{\it et al.\ }{\protect\cite{Edmonds}}, 
the full triangles are the experimental data by Seong 
{\it et al.\ }{\protect\cite{Seong02}}, and 
the dashed line corresponds to a hole concentration equal to that of the 
Mn sites.}
\end{figure}

\begin{figure}[h]
{\centering\resizebox*{2.5in}{!}{\includegraphics*{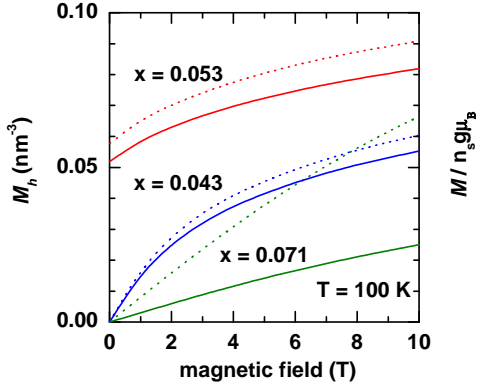}}}
\caption{\label{fig2} 
Magnetizations of the hole ($M_h$; full lines) and Manganese ($M/n_sg\mu B$;
dotted lines) subsystems as functions of the applied magnetic field, for a fixed
temperature of 100 K, and different Mn concentrations.}
\end{figure}

We start with a Hamiltonian for the coupled hole and local moments subsystems in
the form
\begin{eqnarray}
{\cal H}&=&{\cal H}_{{\rm h}}+J_{pd}\sum_{i,I}{\bf S}
_{I}\cdot {\bf s}_{i}\ \delta \left( {\bf r}_{i}-{\bf R}_{I}\right)+\nonumber\\
&+&g_{\mathrm{Mn}}\mu_{\mathrm{B}}\mathbf{H}\cdot\sum_{I}\mathbf{S}_I
-g_{\mathrm{h}}\mu_{\mathrm{B}}\mathbf{H}\cdot\sum_i {\bf s}_{i},
\label{Ham1}
\end{eqnarray}
where the {\em direct} ({\em i.e.,} non--hole-mediated) antiferromagnetic
exchange between Mn spins has been neglected, ${\cal H}_{{\rm h}}$ describes the
hole subsystem, ${\bf S}_{I}$ and ${\bf s}_{i}$ label the localized Mn spins
($S=5/2$) and the hole spins ($s=1/2$), respectively; the second term
corresponds to the Mn-hole exchange interaction, and the last two terms
represent the coupling to the external field $\mathbf{H}$.  Within the spirit of
a mean field approximation, the magnetization of the Mn subsystem is then given
by
\begin{equation}
M=n_{s}g\mu _{{\rm B}}xSB_{S}
\left[ {\frac{S}{2k_{{\rm B}}T}}\left(J_{pd}{\cal M}_{h}+2g\mu_{\rm B}H \right)
\right],
\label{M}
\end{equation}
which must be determined self-consistently with the hole magnetization,
\begin{equation}
{\cal M}_{h}=
A
\left(
      \frac{4 J_{pd}}{a^3}\frac{M}{n_{s}g\mu _{\rm B}}+g\mu _{\rm B}H
\right)
p^{1/3},  
\label{mh}
\end{equation}
where $n_{s}$ is the density of Ga lattice sites, $g=g_h=g_{\rm Mn}=2$,
$\mu_{\rm B}$ is the Bohr magneton, $S=5/2$ is the Mn spin, $B_{S}[\ldots ]$ is
the Brillouin function, $A=(3\pi^2)^{-2/3}(3m^*/2\hbar^2)$, and $a$ is the GaAs
lattice constant; 
the product $m^*J_{pd}^2$ was determined in Ref.\ \cite{dSOC} 
through a fit to experimental data, and we now take $m^*=m_e$, the 
electronic bare mass, which is within the limits recently set by infrared 
spectroscopy \cite{Singley02}. 
From Eq.\ (\ref{mh}), we see that the dependence of $p$ 
with
$x$ influences the magnetic behavior in a fundamental way.  Here we use $p(x)$
as parametrized in terms of a fitting to experimental data on the critical
temperature $T_c$ \cite{dSOC}, and shown as the lower curve in Fig.\ \ref{fig1};
the qualitative agreement of these data with those recently obtained
\cite{Edmonds} by Hall measurements (represented by the squares in Fig.\
\ref{fig1}) indicates that our procedure provides a reliable input to discuss
the magnetic behavior in the presence of an external field.

\begin{figure}
{\centering\resizebox*{2.5in}{!}{\includegraphics*{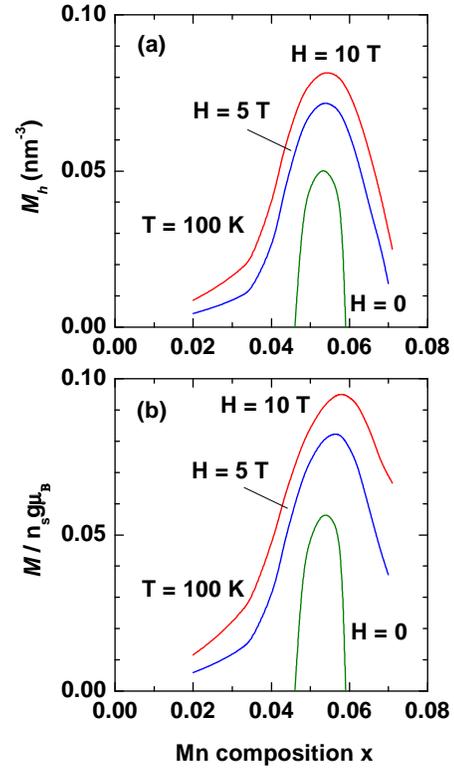}}}
\caption{\label{fig3} 
Magnetizations of the hole (a) and Mn (b) subsystems as functions of the Mn
concentration for different applied magnetic fields, and for a fixed temperature
of 100 K.}
\end{figure}

In Fig.\ \ref{fig2} we show the hole and Mn magnetizations -- obtained
from self-consistent solutions of Eqs.\ (\ref{M}) and (\ref{mh}) -- as
functions of the magnetic field, for a fixed temperature $T=100$ K, and
for three different Mn compositions.  For both $x=0.043$ and $x=0.071$,
the system does not sustain spontaneous magnetic order at this
temperature.  Nonetheless, their magnetic responses to an applied field
are quite distinct.  While for $x=0.043$ the Mn and hole subsystems
display roughly the same susceptibility, for $x=0.071$ the Mn moments
(dotted lines in Fig.\ \ref{fig2}) are more susceptible to the field than
the holes (full lines); this appears as a result of $p(x = 0.071) \ll p(x
= 0.043)$, which, in turn, may be correlated \cite{dSOC} with the fact
that the sample with $x=0.043$ is metallic, and the one with $x=0.071$ is
insulating \cite{Ohno-tc}.

Figure \ref{fig3} displays our results for the Mn and hole magnetizations
at $T=100$ K, for varying Mn compositions and different values of the
external magnetic field.  One notices a sharply peaked structure, with
both maxima and widths increasing with applied magnetic field, as one
would expect.  Again, one finds a difference in the behavior of the Mn and
hole magnetizations:  While the peak position of $M_h$ does not change
with the applied field, the Mn magnetization peaks move slightly towards
larger compositions; it would be interesting to investigate whether this
difference can be detected experimentally, or if it is a mere artifact of
the present approximations.  At any rate, the fact that the magnetization
{\it vs.\ }$x$ curves broaden in the presence of an external field means
that the working window of compositions for spintronics applications is
also broadened.

\begin{figure}[t]
{\centering\resizebox*{2.5in}{!}{\includegraphics*{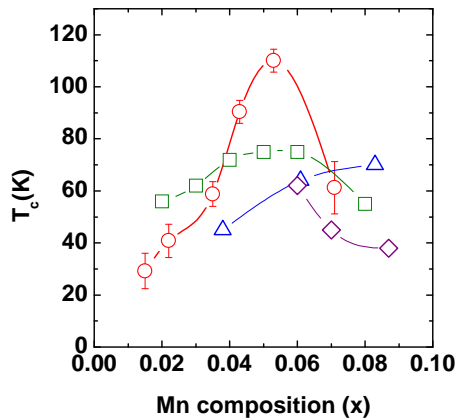}}}
\caption{\label{fig4} 
Experimental $T_c(x)$ in Ga$_{1-x}$Mn$_x$As alloys: circles are data from 
Ohno {\it et al.\ } {\protect\cite{Ohno-tc}}, squares from Edmonds 
{\it et al.\ }{\protect\cite{Edmonds}}, triangles from Seong 
{\it et al.\ }{\protect\cite{Seong02}} and Potashnik 
{\it et al.\ }{\protect\cite{Potashnik}},
and diamonds from van Esch {\it et al.}  {\protect\cite{vanEsch97}}.  All lines
through data points are guides to the eye.}
\end{figure}

At this point it is instructive to resume the discussion of Fig.\ \ref{fig1}.
The theoretical hole concentration $p(x)$ of Fig.\ \ref{fig1}, which was
parametrized via a fitting of the experimental Hall-resistance measurement
\cite{Ohno-tc} at $x=0.053$, $T_c=110$ K, and $p=3.5\times 10^{20}$
cm$^{-3}$ (see cross in the lower curve of Fig.\ \ref{fig1}), is in fair
overall agreement with the more recent Hall experiments of Edmonds {\it et
al.\ }\cite{Edmonds}. By contrast, Raman scattering measurements of the
hole densities, performed by Seong {\it et al.\ } \cite{Seong02} in four
Ga$_{1-x}$Mn$_x$As samples ($x=0,$ 0.038, 0.061, and 0.083), resulted in a
\emph{monotonically increasing} $p(x)$; see the triangles in Fig.\
\ref{fig1}. This monotonic behavior is also in disagreement with the early
results by Matsukura {\it et al.} \cite{Matsukura98}. Seong {\it et al.\
}\cite{Seong02} argue that Raman scattering - unlike standard Hall
measurements - provides an unambiguous and reliable method of determining
the hole density in Ga$_{1-x}$Mn$_x$As systems, and comment that the
difference between their results and those of Ref.\ \cite{Matsukura98}
could be attributed to differences in detailed growth conditions.  
Unfortunately, it seems that the details of growth conditions (even in
as-grown samples) also affects the behavior of $T_c$ with Mn composition:  
In Fig.\ \ref{fig4} we display the critical temperatures as obtained by
several groups \cite{Ohno-tc,Edmonds,Potashnik,Seong02,vanEsch97}:  The
inescapable conclusion is that the measurements of both $p(x)$ and of
$T_c(x)$ in Ga$_{1-x}$Mn$_x$As are strongly sample dependent (or
growth-conditions dependent).  This indicates the need of further detailed
experimental work on Ga$_{1-x}$Mn$_x$As samples in order to assure that
data on $T_c(x)$ and $p(x)$ are reproducible.

Summing up, we have investigated the effects of a magnetic field on the
magnetizations of the hole and Mn subsystems in Ga$_{1-x}$Mn$_x$As
semiconductor compounds. Through a a mean-field approach, we have
established that holes are less susceptible to the magnetic field than Mn
ions at larger dopings; we have also found that the dependence of
the magnetizations with $x$, for a given temperature, presents a sharply
peaked structure, with both maxima and widths increasing with applied
magnetic field, thus indicating that diluted-magnetic-semiconductor
devices would require quality-control of the Mn-doping composition.

\begin{acknowledgments}

The authors are grateful to K.\ W.\ Edmonds and M.\ J.\ Seong for providing
their data in tabular form.  This research was partially supported by Brazilian
Agencies FAPERJ, FAPESP, CNPq, FAEP-UNICAMP, Millenium Institute for
Nanosciences, and Brazilian Network for Nanosciences.

\end{acknowledgments}

\end{document}